\documentclass[conference,10pt]{IEEEtran}
\usepackage{cite}
\usepackage{color}
\usepackage[final]{graphicx}
\usepackage{amsmath,amsfonts}

\newtheorem{theorem}{Theorem}
\def\figref#1{Fig.~\ref{#1}}
\def\secref#1{Sec.~\ref{#1}}

\def\bd{\mathbf d}
\def\bx{\mathbf x}
\def\bf{\mathbf f}
\def\bu{\mathbf u}
\def\bq{\mathbf q}
\def\bp{\mathbf p}
\def\by{\mathbf y}
\def\bs{\mathbf s}
\def\bR{\mathbf R}
\def\bA{\mathbf A}
\def\bW{\mathbf W}
\def\bT{\mathbf T}
\def\bB{\mathbf B}
\def\bnu{\boldsymbol \nu}
\def\bzero{\mathbf 0}
\def\hk{{\hat k}}
\def\cD{{\mathcal D}}
\def\cDin{{\mathcal D}^{(\text{in})}}
\def\cC{{\mathcal C}}
\def\cK{{\mathcal K}}
\def\cN{{\mathcal N}}
\def\cM{{\mathcal M}}
\def\cT{{\mathcal T}}

\def\hk{{\hat k}}
\def\setZ{{\mathbb Z}}
\def\setR{{\mathbb R}}
\def\bH{\mathbf H}
\def\size #1x#2{$#1\times #2$}
\newcommand*{\bydef}{\mathrel{\vcenter{\baselineskip0.5ex \lineskiplimit0pt
   \hbox{\scriptsize.}\hbox{\scriptsize.}}}=}
\newcounter{remark}
\def\remark{\addtocounter{remark}1\emph{Remark \arabic{remark}}: }
\makeatletter
\def\relabel#1{\write\@auxout{\string\newlabel{#1}{{\theremark}{\thepage}}}}
\makeatother
\def\paren#1{\left(#1\right)}
\def\part#1{\medskip\noindent\textsc{#1}}
\def\st{\,\bigr|\,}
\def\thref#1{Theorem~\ref{#1}}

\normalsize
\begin{document}

\title{Multiple-Antenna Interference Channel with Receive Antenna
Joint Processing and Real Interference Alignment}

\author{
  \IEEEauthorblockN{Mahdi Zamanighomi and Zhengdao Wang}
  \IEEEauthorblockA{Dept. of Electrical \& Computer Eng.\\
  Iowa State University\\
  Ames, Iowa, USA\\
  Email: \{mzamani,zhengdao\}@iastate.edu}
}

\maketitle

\begin{abstract}
We consider a constant $K$-user Gaussian interference channel with $M$
antennas at each transmitter and $N$ antennas at each receiver, denoted as a
$(K,M,N)$ channel. Relying on a result on simultaneous Diophantine
approximation, a real interference alignment scheme with joint receive antenna
processing is developed. The scheme is used to provide new proofs for two
previously known results, namely 1) the total degrees of freedom (DoF) of a
$(K, N, N)$ channel is $NK/2$; and 2) the total DoF of a $(K, M, N)$ channel
is at least $KMN/(M+N)$. We also derive the DoF region of the $(K,N,N)$
channel, and an inner bound on the DoF region of the $(K,M,N)$ channel.
\end{abstract}

\section{Introduction} Interference channel is an important model for
multi-user communication systems. In a $K$-user interference channel, the
$k$-th transmitter has a message intended for the $k$-th receiver. At receiver
$k$, the messages from transmitters other than the $k$-th are interference.
Characterizing the capacity region of a general interference channel is an
open problem, although results for some specific cases are known.

To quantify the shape of the capacity region at high signal-to-noise ratio
(SNR), the concept of degrees of freedom (DoF) has been introduced
\cite{mamk06c,caja08}. The DoF of a message is its rate normalized by the
capacity of single-user additive white Gaussian noise channel, as the SNR
tends to infinity.

To achieve the optimal DoF, the concept of interference alignment turns out to
be important \cite{jafa11}. At a receiver, the interference signals from
multiple transmitters are aligned in the signal space, so that the
dimensionality of the interference in the signal space can be minimized.
Therefore, the remaining space is interference free and can be used for the
desired signals. Two commonly used alignment schemes are vector alignment and
real alignment. In real alignment, the concept of linear independence over the
rational numbers replaces the more familiar vector linear independence. And a
Groshev type of theorem is usually used to guarantee the required decoding
performance \cite{mgmk09}.

So far the real alignment schemes have been mainly developed only for scalar
interference channels. For multiple-input multiple output (MIMO) interference
channels, antenna splitting argument has been used in \cite{mgmk09} and
\cite{ghmk10c} to derive the total DoF. In such antenna splitting arguments,
no cooperation is employed either at the transmitter side or at the receiver
side.

In this paper, we consider a constant $K$-user Gaussian interference channel
with $M$ antennas at each transmitter and $N$ antennas at each receiver,
denoted as a \emph{$(K,M,N)$ channel}. We develop a real alignment scheme for
MIMO interference channel that employs joint receive antenna processing.
Relying on the recent results on simultaneous Diophantine approximation, we
are able to obtain new proofs of two previously known results, and derive two
new results on the DoF region; see \secref{sec.main}.

\section{System model}

Notation: $K$, $D$, $D'$, and $N$ are integers and $\cK=\{1, \ldots, K\}$,
$\cN=\{1, \ldots, N\}$. We use $k$ and $\hk$ as transmitter indices, and $j$
as receiver indices. Superscripts $t$ and $r$ are used for transmitter and
receiver antenna indices. The set of integers and real numbers are denoted as
$\setZ$ and $\setR$, respectively. The set of non-negative real numbers is
denoted as $\setR_+$. Letter $i$ and $l$ are used as the indices of directions
and streams, respectively. Vectors and matrices are indicated by bold symbols.
We use $\|\bx\|$ to denote infinity norm of $\bx$, $(\cdot)^*$ matrix
transpose, and $\otimes$ the Kronecker product of two matrices.

Consider a multiple-antenna $K$-user real Gaussian interference channel with
$M$ antennas at each transmitter and $N$ antennas at each receiver. At each
time, each transmitter, say transmitter $k$, sends a vector signal
$\bx_k\in\setR^{M}$ intended for receiver $k$. The channel from transmitter
$k$ to receiver $j$ is represented as a matrix
\begin{equation}
\bH_{j,k}\bydef [h_{j,k,r,t}]_{r=1,t=1}^{N,M}
\end{equation}
where $k\in\cK$, $j\in\cK$, and $\bH_{j,k}\in \setR^{N\times M}$. It is
assumed that the channel is constant during all transmissions. Each
transmitter is subjected to a power constraint $P$. The received signal at
receiver $j$ can be expressed as
\begin{equation}
\by_{j} = \sum_{k\in \cK} \bH_{j,k} \bx_{k} +
  \bnu_{j}, \quad \forall j\in \cK
\end{equation}
where $\{\bnu_{j}|j\in \cK\}$ is the set of independent Gaussian additive
noises with real, zero mean, independent, and unit variance entries. Let $\bH$
denote the $(KN)\times (KM)$ block matrix, whose $(j,k)$th block of size \size
NxM is the matrix $\bH_{j,k}$. The matrix $\bH$ includes all the channel
coefficients. For the interference channel $\bH$, the \emph{capacity region}
$\cC(P,K,\bH)$ is defined in the usual sense: It contains rate tuples
$\bR_K(P)=[R_1(P), R_2(P), \ldots, R_K(P)]$ such that reliable transmission
from transmitter $k$ to receiver $k$ is possible at rate $R_k$ for all $k\in
\cK$ simultaneously, under the given power constraint $P$. Reliable
transmissions mean that the probability of error can be made arbitrarily small
by increasing the encoding block length while keeping the rates and power
fixed.

A DoF vector $\bd=(d_1, d_2, \ldots, d_K)$ is said to be \emph{achievable} if
for any large enough $P$, the rates $R_i=0.5\log(P) d_i$, $i=1,2,\ldots, K$,
are simultaneously achievable by all $K$ users, namely
\(0.5\log(P)\cdot \bd \in \cC(P,K,\bH)\), for $P$ large enough.
The \emph{DoF region} for a given channel $\bH$, $\cD(K,\bH)$, is the closure
of the set of all achievable DoF vectors. The DoF region $\cD(K,M,N)$ is the
largest possible region such that $\cD(K,M,N)\subset \cD(K,\bH)$ for almost
all $\bH$ in the Lebesgue sense. The \emph{total DoF of the $K$-user
interference channel $\bH$} is defined as
\[
d(K,\bH)=\max_{\bd\in \cD(K,\bH)} \sum_{k=1}^K d_k.
\]
The \emph{total DoF $d(K,M,N)$} is defined as the largest possible real number
$\mu$ such that for almost all (in the Lebesgue sense) real channel matrices
$\bH$ of size $(KN)\times (KM)$, $d(K,\bH)\ge \mu$.

\section{Main Results} \label{sec.main}

The following two theorems have been proved before, in \cite{mgmk09} and
\cite{ghmk10c}, respectively:
\begin{theorem} \label{th.KNN}
$d(K,N,N)=\frac{NK}{2}$.
\end{theorem}

\begin{theorem} \label{th.KMN}
$d(K,M,N)\ge \frac{MN}{M+N}K$.
\end{theorem}

The main contributions of the paper are 1) providing alternative proofs of the
above two theorems, and 2) prove the following theorems.

\begin{theorem}\label{th.region.KNN}
The DoF region of a $(K,N,N)$ interference channel is the following
\[
\cD(K,N,N)=\{\bd\in \setR_+^{K\times 1}\st d_k + \max_{\hat k\ne k} d_{\hat k}
\le N, \forall k\in \cK \}.
\]
\end{theorem}

\begin{theorem}\label{th.region.KMN}
The DoF region of a $(K,M,N)$ interference channel satisfies
$\cD(K,N,N)\supset \cDin$ where
\begin{multline*}
\cDin\bydef \{\bd\in \setR_+^{K\times 1}\st M d_k + N\max_{\hat k\ne k}
d_{\hat k} \le MN, \\ \forall k\in \cK\}.
\end{multline*}
\end{theorem}

\remark The DoF region of $K$-user time-varying interference channel with $N$
antennas at each node has been obtained before in \cite{krwy12i}. The fact the
DoF region of a $(K, N, N)$ channel in \thref{th.region.KNN} is the same as
that of a time-varying MIMO interference channel with the same number of
antennas indicates that the DoF region for this channel is an inherent spatial
property of the channel that is separate from the time or frequency diversity.

\remark \thref{th.region.KNN} follows from \thref{th.region.KMN} by setting
$M=N$, and the Multiple Access Channel (MAC) outer bound
\cite[Sec.~14.3]{coth91}.

\remark \thref{th.KMN} follows from \thref{th.region.KMN} by setting
$d_k=MN/(M+N)$, $\forall k\in \cK$.

\remark \thref{th.KNN} follows from \thref{th.KMN} by setting $M=N$ and the
outer bound for $K$-user interference channel that has been obtained before in
\cite{caja08}. \relabel{re.KNN}

From the above remarks, it is only necessary to prove \thref{th.region.KMN}.
However, we will first prove the achievability of \thref{th.KNN} in the next
section, which serves to introduce the joint antenna processing at the
receivers, and the application of the result in simultaneous Diophantine
approximation on manifolds. \thref{th.region.KMN} will be proved in
\secref{sec.region.KMN}.

\begin{figure}[tbp]
\centering
\includegraphics[width=.9\linewidth]{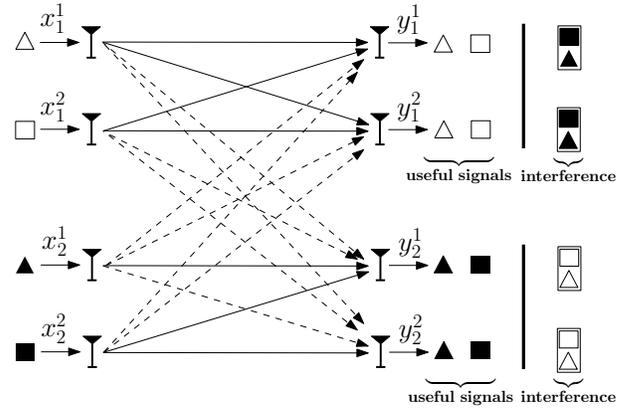}
\caption{2-user Gaussian interference channel with 2 antennas at each
transmitter receiver}
\label{fig.1}
\end{figure}

\section{Achievability Proof for \thref{th.KNN}}

One important technique for proving achievability result is the real
interference alignment \cite{mgmk09} which seeks to align the dimensions of
interferences so that more free dimensions can be available for intended
signals. The dimensions (also named directions) are represented as real
numbers that are rationally independent (cf.~Appendix A).

We will denote set of directions, a specific direction, and vector of
directions using $\cT$, $T$, and $\bT$ respectively.

\part{encoding}: Transmitter $k$ sends a vector message $\bx_k =
{(x^1_k,\ldots,x^N_k)}^*$ where $x^t_k$, $\forall t\in\cN$ is the signal
emitted by antenna $t$ at transmitter $k$. The signal $x^t_k$ is generated
using transmit directions in a set $\cT=\{T_i\in\setR|1\leq i\leq D\}$ as
follows

\begin{equation}
x^t_k = \bT \bs^t_k
\end{equation}
where
\begin{equation}
\bT \bydef (T_1,\ldots ,T_D), \quad
\bs^t_k \bydef {(s^t_{k1},\ldots , s^t_{kD})}^*
\end{equation}
and $\forall 1 \leq i \leq D$,
\begin{equation}\label{alpha}
s^t_{ki} \in \{\lambda q \st q\in\setZ, -Q\leq q\leq Q\}.
\end{equation}
The parameters $Q$ and $\lambda$ will be designed to satisfy the rate and
power constraints.

\part{Alignment Design}: We design transmit directions in such a way that at
any receiver antenna, each useful signal occupies a set of directions that are
rationally independent of interference directions (cf.~Appendix A).

To make it more clear, consider \figref{fig.1}. $x^1_1$ and $x^2_1$ are shown
by white triangle and square. In a similar fashion, $x^1_2$ and $x^2_2$ are
indicated with black triangle and square. We are interested in such transmit
directions that at each receiver antenna the interferences, for instance black
triangle and square at receiver $1$, are aligned while the useful messages,
white triangle and square, occupy different set of directions.

\part{transmit directions}: Our scheme requires all transmitter antennas to
only contain directions of the following form
\begin{equation}\label{eq.T}
T = \prod_{j\in \cK} \prod_{k\in\cK, k\ne j} \prod_{r\in\cN}\prod_{t\in\cN}
  \paren{h_{j,k,r,t}}^{\alpha_{j,k,r,t}}
\end{equation}
where
\begin{equation}
0\leq \alpha_{j,k,r,t}\leq n-1,
\end{equation}
$\forall j\in \cK$, $k\in\cK$, $k\ne j$, $r\in\cN$, $t\in\cN$. It is easy to
see that the total number directions is
\begin{equation}
D = n^{K(K-1)N^2}.
\end{equation}
We also assume that directions in $\cT$ are indexed from $1$ to $D$. The exact
indexing order is not important here.

\part{alignment analysis}: Our design proposes that at each antenna of
receiver $j$, $j\in\cK$, the set of messages $\{x_k^t | k\in\cK, k\neq j,
t\in\cN\}$ are aligned. To verify, consider all $x_k^t$, $k\neq j$ that are
generated in directions of set $\cT$. These symbols are interpreted as the
interferences for receiver $j$. Let
\begin{equation}
D' = {(n+1)}^{K(K-1)N^2}.
\end{equation}
and define a set $\cT'=\{T'_i\in\setR |1\leq i\leq D'\}$ such that all $T'_i$
are in from of $T$ as in \eqref{eq.T} but with a small change as follows
\begin{equation}\label{eq.T'}
0\leq \alpha_{j,k,r,t}\leq n.
\end{equation}
Clearly, all $x_k^t$, $k\neq j$ arrive at antenna $r$ of receiver $j$ in the
directions of $\{\paren{h_{j,k,r,t}}T | k\in\cK, k\neq j, t\in\cN, T\in\cT\}$
which is a subset of $\cT'$.

This confirms that at each antenna of any receiver, all the interferences only
contain the directions from $\mathcal{T'}$. These interference directions can
be described by a vector
\begin{equation}
\bT'  \bydef (T'_1,\ldots , T'_{D'}).
\end{equation}

\part{decoding scheme}: In this part, we first rewrite the received signals.
Then, we prove the achievability part of \thref{th.KNN} using joint antenna
processing.

The received signal at receiver $j$ is represented by
\begin{equation}
\by_j=\bH_{j,j}\bx_j + \sum_{k\in\cK,k\neq j}{\bH_{j,k}\bx_k} + \bnu_{j}.
\end{equation}

Let us define
\begin{equation}
\bB \bydef \left(\begin{matrix}
  \bT &  \bzero  & \ldots & \bzero \cr
  \bzero   &  \bT & \ldots & \bzero \cr
  \vdots & \vdots &\ddots & \vdots\cr
  \bzero   &   \bzero &\ldots & \bT
  \end{matrix}\right)
\end{equation}
and
\begin{equation}
\bs_k \bydef \left(\begin{matrix}
   \bs^1_k \cr
   \bs^2_k \cr
   \vdots \cr
   \bs^N_k
   \end{matrix}\right), \quad
\bu_k=\frac{\bs_k}{\lambda},
\end{equation}
such that $\bB$ is a $N\times ND$ matrix with $(N-1)D$ zeros at each row.
Using above definitions, $\by_j$ can be rewritten as
\begin{equation}
\by_j=\lambda\paren{ \bH_{j,j}\bB\bu_j+ \sum_{k\in\cK,k\neq
j}{\bH_{j,k}\bB\bu_k}} + \bnu_{j}.
\end{equation}
The elements of $\bu_k$ are integers between $-Q$ and $Q$, cf.~\eqref{alpha}.

We rewrite
\begin{multline}\label{reshape1}
\bH_{j,j}\bB\bu_j = \paren{\bH_{j,j}\otimes \bT} \bu_j = \\
				\left(\begin{matrix}
  h_{j,j,1,1}\bT &  h_{j,j,1,2}\bT  & \ldots & h_{j,j,1,N}\bT \cr
  h_{j,j,2,1}\bT   &  h_{j,j,2,2}\bT & \ldots & h_{j,j,2,N}\bT \cr
  \vdots & \vdots &\ddots & \vdots\cr
  h_{j,j,N,1}\bT & h_{j,j,N,2}\bT & \ldots & h_{j,j,N,N}\bT
  \end{matrix}\right)\bu_j \bydef
  \left(\begin{matrix}
  \bT^1_j\cr
  \bT^2_j\cr
  \vdots\cr
  \bT^N_j
  \end{matrix}\right)\bu_j
\end{multline}
where $\forall r\in\cN$, $\bT^r_j$ is the $r^{\text{th}}$ row of
$\bH_{j,j}\bB$. Also,
\begin{multline}\label{reshape2}
\sum_{k\in\cK,k\neq j}{\bH_{j,k}\bB\bu_k} = \sum_{k\in\cK,k\neq
j}{\paren{\bH_{j,k}\otimes\bT}\bu_k} = \\
				 \left(\begin{matrix}
  \sum_{k\in\cK, k\neq j}{\sum_{t\in \cN}{\paren{ h_{j,k,1,t}\bT\bu^t_k}}}\cr
  \sum_{k\in\cK, k\neq j}{\sum_{t\in \cN}{\paren{ h_{j,k,2,t}\bT\bu^t_k}}}\cr
  \vdots\cr
  \sum_{k\in\cK, k\neq j}{\sum_{t\in\cN}{\paren{ h_{j,k,N,t}\bT\bu^t_k}}}
  \end{matrix}\right) { }^{(a)}_{\ =}
  \left(\begin{matrix}
  \bT'\bu'^1_j\cr
  \bT'\bu'^2_j\cr
  \vdots\cr
  \bT'\bu'^N_j
  \end{matrix}\right)
\end{multline}
where $\forall r\in\cN$, $\bu'^r_j$ is a column vector with $D'$ integer
elements, and $(a)$ follows since the set $\cT'$ contains all directions of
the form $\paren{h_{j,k,r,t}}T$ where $k\neq j$; cf.~the definition of $\cT'$.

Considering (\ref{reshape1}) and (\ref{reshape2}), we are able to equivalently
denote $\by_j$ as
\begin{equation}\label{zeros}
\by_j = \lambda \left(\begin{matrix}
   \bT^1_j &\bT' &  \bzero  & \ldots & \bzero \cr
  \bT^2_j &\bzero   &  \bT' & \ldots & \bzero \cr
  \vdots &\vdots & \vdots &\ddots & \vdots\cr
  \bT^N_j &\bzero   &   \bzero &\ldots & \bT'
  \end{matrix}\right)
  \left(\begin{matrix}
  \bu_j\cr
  \bu'^1_j\cr
  \vdots\cr
  \bu'^N_j
  \end{matrix}\right)
  + \bnu_{j}.
\end{equation}

We finally left multiply $\by_j$ by an $N\times N$ weighting matrix
\begin{equation}
\bW= \left(\begin{matrix}
  1 &  \gamma_{12}& \ldots & \gamma_{1N}  \cr
  \gamma_{21}  & 1 & \ldots & \gamma_{2N} \cr
  \vdots & \vdots & \ddots & \vdots \cr
  \gamma_{N1}& \gamma_{N2} &\ldots & 1
  \end{matrix}\right)
\end{equation}
such that all indexed $\gamma$ are randomly, independently, and uniformly
chosen from interval $[\frac{1}{2}, 1]$. This process causes the zeros in
(\ref{zeros}) to be filled by non-zero directions.

After multiplying $\bW$, the noiseless received constellation belongs to a
lattice generated by the $N\times N(D+D')$ matrix
\begin{equation}
 \bA = \bW \left(\begin{matrix}
   \bT^1_j &\bT' &  \bzero  & \ldots & \bzero \cr
  \bT^2_j &\bzero   &  \bT' & \ldots & \bzero \cr
  \vdots &\vdots & \vdots &\ddots & \vdots\cr
  \bT^N_j &\bzero   &   \bzero &\ldots & \bT'
  \end{matrix}\right).
\end{equation}

The above matrix has a significant property that allows us to use
\thref{Th.two} (cf.~Appendix C). \thref{Th.two} requires each row of $\bA$ to
be a nondegenerate map from a subset of channel coefficients to
$\setR^{N(D+D')}$. The nondegeneracy is established because (cf.~Appendix B):
\begin{enumerate}
\item all elements of $\bT'$ and $\bT^t_j$, $\forall t\in\cN$ are analytic
functions of the channel coefficients;
\item all the directions in $\bT'$ and $\bT^t_j$, $\forall t\in\cN$ together
with 1 are linearly independent over $\setR$ ;
\item all indexed $\gamma$ in $\bW$ have been chosen randomly and
independently.
\end{enumerate}
Hence, using \thref{Th.two}, the set of $\bH$ such that there exist infinitely
many
\begin{equation}
\bq = \left(\begin{matrix}\nonumber
  \bu_j\cr
  \bu'^1_j\cr
  \vdots\cr
  \bu'^N_j
  \end{matrix}
  \right)\in\setZ^{N(D+D')}
\end{equation}
with
\begin{equation}
\|\bA\bq\| < \|\bq\|^{-(D+D')-\epsilon} \quad \text{for  } \epsilon>0
\end{equation}
has zero Lebesgue measure. In other words, for almost all $\bH$, $\|\bA\bq\| >
\|\bq\|^{-(D+D')-\epsilon}$ holds for all $\bq\in\setZ^{N(D+D')}$ except for
finite number of them. By the construction of $\bA$, all elements in each row
of $\bA$ are rationally independent with probability one, which means that
$\bA\bq\neq \bzero$ unless $\bq=\bzero$. Therefore, almost surely for any
fixed channel (hence fixed $\bA$), there is a positive constant $\beta$ such
that $\|\bA\bq\| >\beta \|\bq\|^{-(D+D')-\epsilon}$ holds for all integer
$\bq\neq\bzero$. Since
\[
\|\bq\|\le (K-1)NQ,
\]
the distance between any two points of received constellation (without
considering noise) is lower bounded by
\begin{equation}\label{result1}
\beta \lambda \bigl(\,(K-1)NQ\,\bigr)^{-(D+D')-\epsilon}.
\end{equation}

\remark The noiseless received signal belongs to a constellation of the form
\begin{equation}
\by=\lambda\bA\bar{\bq}
\end{equation}
where $\bar{\bq}$ is an integer vector. Then, the hard decision maps the
received signal to the nearest point in the constellation. Note that the hard
decoder employs all $N$ antennas of receiver $j$ to detect signals emitted by
intended transmitter. In other words, our decoding scheme is based on
multi-antenna joint processing.

We now design the parameters $\lambda$ and $Q$. With reference to
\cite{mgmk09}, if we choose
\begin{equation}
\lambda =\zeta \frac{P^{\frac{1}{2}}}{Q},
\end{equation}
then the power constraint is satisfied ($\zeta$ is a constant here). Moreover,
similar to \cite{mgmk09}, we choose
\begin{equation}\label{result2}
Q = P^{\frac{1-\epsilon}{2(D+D'+1+\epsilon)}} \quad \text{for  } \epsilon\in(0,1)
\end{equation}
assuring that the DoF per direction is $\frac{1-\epsilon}{D+D'+1+\epsilon}$.
Since, we are allowed to arbitrarily choose $\epsilon$ within $(0,1)$,
$\frac{1}{D+D'+1}$ is also achievable.

Using \eqref{result1}--\eqref{result2} and the performance analysis described
by \cite{mgmk09}, the hard decoding error probability of received
constellation goes to zero as $P\rightarrow \infty$ and the total achievable
DoF for almost all channel coefficients in the Lebesgue sense is
\begin{equation}
\frac{NKD}{D+D'+1} =\frac{NKn^{K(K-1)N^2}}{n^{K(K-1)N^2}+{(n+1)}^{K(K-1)N^2}+1}
\end{equation}
and as $n$ increases, the total DoF goes to $\frac{NK}{2}$ which meets the
outer bound \cite{caja08}.

\section{Proof of \thref{th.region.KMN}}\label{sec.region.KMN} Notation:
Unless otherwise stated, all the assumptions and definitions are still the
same.

Consider the case where the number of transmitter and receiver antennas are
not equal. This can be termed the $K$-user MIMO interference channel with $M$
antennas at each transmitter and $N$ antennas at each receiver. Hence, for all
$j\in\cK$ and $K\in\cK$, $\bH_{j,k}$ is a $N\times M$ matrix. We prove that
for any $\bd\in\cDin$, $\bd$ is achievable.

Under the rational assumption, it is possible to find an integer $\rho$ such
that $\forall k\in\cK$, ${\bar{d}}_k=\rho \frac{d_k}{M}$ where ${\bar{d}}_k$
is a non-negative integer. The signal $x_k^t$ is divided into ${\bar{d}}_k$
streams. For stream $l$, $l\in
\{1,\ldots,\displaystyle\max_{k\in\cK}{\bar{d}_k}\}$, we use directions
$\{T_{l1},\ldots, T_{lD}\}$ of the following form
\begin{equation}
T_l = \prod_{j\in \cK} \prod_{k\in\cK, k\ne j} \prod_{r\in\cN}\prod_{t\in\cN}
  \paren{h_{j,k,r,t}\delta_l}^{\alpha_{j,k,r,t}}
\end{equation}
where $0 \leq \alpha_{j,k,r,t} \leq n-1$ and $\delta_l$ is a design parameter
that is chosen randomly, independently, and uniformly from the interval
$[\frac{1}{2},1]$. Let $\bT_l \bydef (T_{l1}, \ldots, T_{lD})$. Note that, at
any antenna of transmitter $k$, the constants $\{\delta_l\}$ cause the streams
to be placed in $\bar{d}_k$ different sets of directions. The alignment scheme
is the same as before, considering the fact that at each antenna of receiver
$j$, the useful streams occupy $M\bar{d}_j$ separate sets of directions. The
interferences are also aligned at most in $\displaystyle\max_{k\in\cK, k\neq
j}{\bar{d}_k}$ sets of directions independent from useful directions.

By design, $x_k^t$ is emitted in the following form
\begin{equation}
x_k^t = \sum_{l=1}^{\bar{d}_k}{\delta_l}\sum_{i=1}^{D}{T_{li}s_{kli}^t} =
\bT_k\bs^t_k
\end{equation}
where
\begin{equation}
\bT_k\bydef(\delta_1 \bT_{1},\ldots,\delta_{\bar{d}_k}\bT_{\delta_{\bar{d}_k}}),
\end{equation}
\begin{equation}
\bs^t_k \bydef {(s^t_{k11},\ldots , s^t_{k\bar{d}_kD})}^*,
\end{equation}
and all $s_{kli}^t$ belong to the set defined in (\ref{alpha}).

Pursuing the same steps of the previous section for receiver $j$, $\bB$
becomes a $M\times MD\bar{d}_j$ matrix and $\bA$ will have $N$ rows and
$MD{\bar{d}}_j+ND'\displaystyle\max_{k\in\cK, k\neq j}{{\bar{d}}_k}$ columns.
The total number directions $G_j$ of both useful signals and the interferences
at receiver $j$ satisfies
\begin{equation}
G_j \leq MD{\bar{d}}_j  + ND'{\displaystyle\max_{k\in\cK, k\neq j}{{\bar{d}}_k}}.
\end{equation}
For any DoF points in $\cDin$ satisfying \thref{th.region.KMN}, we have
\begin{equation}
G_j \leq \paren{{M{\bar{d}}_j + N{\displaystyle\max_{k\in\cK, k\neq j}{{\bar{d}}_k}}}}D' \leq \frac{\rho}{M}NMD' = \rho ND'
\end{equation}
and as $n$ increases, the DoF of the signal $x_j$ intended for receiver $j$,
$\forall j\in\cK$ can be arbitrarily close to
\begin{multline}
\lim_{n\to\infty}{MD\bar{d}_j\frac{N}{\rho ND'}}= \\
\lim_{n\to\infty}\frac{M}{\rho}\frac{\bar{d}_jn^{K(K-1)N^2}}{{(n+1)}^{K(K-1)N^2}}
= \frac{M}{\rho}\bar{d}_j = d_j
\end{multline}
where $\frac{N}{\rho ND'}$ is the DoF per direction for large $D'$. This
proves \thref{th.region.KMN}.

As a special case, it is easy to see when all $d_k$ are equal, the total
achievable DoF is $\frac{MN}{M+N}K$. Moreover, when $M=N$, the achievable DoF
region meets the outer bound \cite{caja08}.

\section{Conclusions and Future Works}

We developed a new real interference alignment scheme for multiple-antenna
interference channel that employs joint receiver antenna processing. The
scheme utilized a result on simultaneous Diophantine approximation and aligned
all interferences at each receive antenna. We were able to provide new proofs
for two existing results on the total DoF of multiple antenna interference
channels (\thref{th.KNN} and \thref{th.KMN}) and drive two new DoF region
results (\thref{th.region.KNN} and \thref{th.region.KMN}).

It is desired to extend the result of the paper to a multiple-antenna
interference network with $K$ transmitters and $J$ receivers where each
transmitter sends an arbitrary number of messages, and each receiver may be
interested in an arbitrary subset of the transmitted messages. This channel is
known as wireless X network with general message demands.

\medskip \noindent \emph{Acknowledgment}: The authors thank V. Beresnevich for
discussion on the convergence problem of Diophantine approximation on
manifolds and directing us to reference \cite{klmw09}.

\appendix
\subsection{Definitions of Independence}

A set of real numbers is \emph{rationally independent} if none of the elements
of the set can be written as a linear combination of the other elements with
rational coefficients.

A set of functions are \emph{linearly independent over $\setR$} if none of the
functions can be represented by a linear combination of the other functions
with real coefficients.

\subsection{Nondegenerate Manifolds \cite{klma98}} Consider a $d$-dimensional
sub-manifold $\cM=\{\bf(\bx) | \bx\in U\}$ of $\setR^n$, where
$U\subset\setR^d$ is an open set and $\bf= \paren{f_1,\ldots,f_n}$ is a $C^k$
embedding of $U$ to $\setR^n$. For $l\leq k$, $\bf(\bx)$ is an
\emph{$l$-nondegenerate} point of $\cM$ if partial derivatives of $\bf$ at
$\bx$ of order up to $l$ span the space $\setR^n$. The function $\bf$ at $\bx$
is \emph{nondegenerate} when it is $l$-nondegenerate at $\bx$ for some $l$.

If the functions $f_1,\ldots,f_n$ are analytic, and $1,f_1,\ldots,f_n$ are
linearly independent over $\setR$ in a domain $U$, all points of $\cM=\bf(U)$
are nondegenerate.

\subsection{Diophantine approximation for systems of linear forms \cite{klmw09}} Consider a $m\times n$ real matrix $\bA$ and
$\bq\in\setZ^n$. The theory of simultaneous Diophantine approximation tries to
figure out how small the distance from $\bA\bq$ to $\setZ^m$ could be. This
can be viewed as a generalization of estimating real numbers by rationals
\cite{klmw09}.

\begin{theorem}\label{Th.two}
\cite{klmw09} : Let $\bf_i$, $i=1,\ldots,m$ be a nondegenerate map from an
open set $U_i \subset\setR^{d_i}$ to $\setR^n$ and
\begin{equation} \nonumber
F : U_1 \times \ldots \times U_m \rightarrow \cM_{m,n}, \quad
(\bx_1,\ldots,\bx_m)\longmapsto \left( \begin{array}{ccc}
\bf_1(\bx_1)  \\
\vdots \\
\bf_m(\bx_m)  \end{array} \right)
\end{equation}
where $\cM_{m,n}$ denotes the space of $m\times n$ real matrices.

Then, for $\epsilon > 0$, the set of $(\bx_1,\ldots,\bx_m)$ such that for
\begin{equation}
\bA = \left( \begin{array}{ccc}
\bf_1(\bx_1)  \\
\vdots \\
\bf_m(\bx_m)  \end{array} \right)
\end{equation}
there exist infinitely many $\bq \in \setZ^n$ with
\begin{equation}
\|\bA\bq-\bp\| < \|\bq\|^{\frac{-n}{m}-\epsilon} \quad \text{for some} \ \bp\in \setZ^m
\end{equation}
has zero Lebesgue measure on $U_1 \times \ldots \times U_m$.
\end{theorem}

\bibliography{refs}
\bibliographystyle{IEEE-unsorted}

\end{document}